\documentclass[twocolumn]{aastex63}
\usepackage{amsmath,amssymb,amstext}
\usepackage{gensymb}

\shorttitle{Stop, Drop, Don't Bin!}
\shortauthors{May \& Stevenson}

\begin{document}

\title{Introducing a New {\em Spitzer} Master BLISS Map to Remove the Instrument-Systematic -- Phase-Curve-Parameter Degeneracy, as Demonstrated by a Reanalysis of the 4.5$\micron$ WASP-43b Phase Curve}

\correspondingauthor{E. M. May}
\email{Erin.May@jhuapl.edu}

\author[0000-0002-2739-1465]{E. M. May}
\affiliation{Johns Hopkins APL, 11100 Johns Hopkins Rd, Laurel, MD 20723, USA}

\author[0000-0002-7352-7941]{K. B. Stevenson}
\affiliation{Johns Hopkins APL, 11100 Johns Hopkins Rd, Laurel, MD 20723, USA}

\begin{abstract}
    While {\em Spitzer} IRAC systematics are generally well understood, each data set can provide its own challenges that continue to teach us about the underlying functional form of these systematics. Multiple groups have analyzed the phase curves of WASP-43b with varying detrending techniques, each obtaining different results. In this work, we take another look at WASP-43b while exploring the degenerate relation between Bilinearly-Interpolated Subpixel Sensitivity (BLISS) mapping, point response function full width at half-maximum (PRF-FWHM) detrending, and phase curve parameters. We find that there is a strong correlation between the detrending parameters in the two models, and best fit phase curve amplitudes vary strongly when the data are temporally binned. To remove this degeneracy, we present a new Gaussian centroided intrapixel sensitivity map (hereafter fixed sensitivity map), generated using 3,712,830 exposures spanning 5 years, for a variety of aperture sizes at 4.5$\micron$. We find evidence for time variability in the sensitivity at 3.6$\micron$ and do not generate a visit-independent map for this channel. With the fixed 4.5 $\micron$ intrapixel sensitivity map, the best fits for WASP-43b no longer vary strongly with bin size and PRF-FWHM detrending is no longer required to remove correlated noise. For data sets that do not fall completely within the sweet spot, temporal binning should not be used in the analysis of {\em Spitzer} phase curves. We confirm night side emission for WASP-43b with a disk integrated nightside temperature of 806 $\pm$ 48 K at 4.5$\micron$. The 4.5$\micron$ maps are available at \url{github.com/kevin218/POET}. 
\end{abstract}

\section{Introduction}
\par Until the launch of the James Webb Space Telescope, the now retired {\em Spitzer} IRAC \citep{Fazio2004} remains our only source of infrared observations of exoplanet atmospheres. While {\em Spitzer} IRAC \added{photometry} is dominated by instrumental systematics, they are generally well understood and modeled \citep{Charbonneau2005, Agol2010, Knutson2011, Ingalls2012, Stevenson2012, Lewis2013, Deming2015, Mendonca2018, Morello2019} with the primary systematic in both the 3.6 and 4.5 $\micron$ channels a result of intra-pixel sensitivities variations on the detector. \cite{Stevenson2012} introduced Bilinearly-Interpolated Subpixel Sensitivity (BLISS) mapping to fit for these subpixel variations in each observation independently. However, because {\em Spitzer} exoplanet phase-curve observations are commonly split between multiple astronomical observation requests (AORs), occasionally the centroids between multiple AORs for a single phase curve do not overlap and the BLISS map becomes difficult to accurately constrain due to its flexibility through this `self-calibration' fitting. Due to the length of exoplanet phase curve observations, significant centroid drifts with time can further affect the performance of BLISS mapping because of the additional correlation of centroid position with time.
\par In addition to BLISS mapping, \cite{Lanotte2014, Demory2016a, Demory2016b, Gillon2012, Mendonca2018} also used point response function (PRF) full width at half-maximum (FWHM) detrending to remove a further systematic due to the shape of the PRF changing as the centroid drifts away from the center of the pixel. Joint subpixel sensitivity mapping and PRF-FWHM detrending was applied to WASP-43b by \cite{Mendonca2018}, KELT-9b by \cite{Mansfield2020}, and Qatar-1b by \cite{Keating2020}.
\par Previous analyses of the WASP-43b 3.6 and 4.5 $\micron$ phase curves have produced varying results depending on the exact reduction methods used. \cite{Stevenson2017} used only BLISS mapping and finds no night side emission, while \cite{Mendonca2018} binned the data in time and used their own subpixel mapping routine and PRF-FWHM decorrelation to find night-side emission with a lower signal-to-noise ratio (SNR) than \cite{Stevenson2017}. Recently, \cite{Morello2019} tested both a wavelet indepdendent component analysis (ICA) \citep{Morello2016} and a time ICA \citep{Morello2015} technique to identify the important instrumental effects, while also binning the data, with a final result detecting night-side emission between the levels of \cite{Mendonca2018} and \cite{Stevenson2017}. We discus these three different analyses further in Section \ref{W43}.
\par In this article, we present a significant degeneracy between subpixel mapping routines and PRF-FWHM detrending when both are fit for simultaneously, particularly when binning the data in time. To provide a solution for this degeneracy and the occasional lack of overlapping AORs, a fixed intrapixel sensitivity map can be used to better constrain the subpixel variations. A fixed intrapixel map has previously been calculated for IRAC at both 3.6 and 4.5 $\micron$ by \cite{Ingalls2012} \added{\citep[updated in][]{Ingalls2018}} using a center of light centroiding algorithm (written in IDL)\replaced{ that does not perform as well as other centroiding techniques \citep{Lust2014}.}{. While \cite{Lust2014} find that center of light does not perform as well as other centroiding routines, \cite{Ingalls2020} find that center of light is the more accurate method, with the exception of the 8.0$\micron$ channel.}  \added{In this work we use} the POET \citep[Photometry for Orbits, Eclipses, and Transits][]{Campo2011,Stevenson2012,Cubillos2013}) pipeline which applies a Gaussian centroiding technique\replaced{, which was also demonstrated in \cite{Lust2014} to have significant differences in measured centroids as compared to center of light.}{. \cite{Lust2014} demonstrated that Gaussian centroiding and center of light centroiding result in significant difference in the location of the identified center.} Due to these centroiding differences, the \citeauthor{Ingalls2012} maps are not applicable to our reduction methods. In this work, we generate new IRAC intrapixel sensitivity maps using 2D Gaussian centroiding at both 3.6 and 4.5 $\micron$ for a variety of aperture sizes. We present strong evidence for time variability in the 3.6$\micron$ channel's sensitivity, as well as a time-invariable fixed sensitivity map for the 4.5$\micron$ channel that covers the entire `sweet spot'. 
\par In Section \ref{degeneracy} we discuss the degeneracies introduced by using both BLISS mapping and PRF decorrelation as applied to WASP-43b. Section \ref{mastermap} outlines the calibration data and process used to generate our new fixed sensitivity BLISS map, as well as discusses the time variability we see at 3.6$\micron$. In Section \ref{W43} we present our new analysis of the WASP-43b's 4.5$\micron$ phase curve with this fixed intra-pixel sensitivity map. Finally, in Section \ref{conclusions} we summarize this work.

\section{{\em Spitzer} Observations of WASP-43b and Data Reduction}
\label{methods}
\par \added{In this work we consider the WASP-43b {\em Spitzer} IRAC phase curves (Programs 10169 and 11001, PI: Kevin Stevenson), observed during the warm mission in 2014 and 2015. For details on the observational setup for these programs see Section 2 of \cite{Stevenson2017}. Each phase curve is reduced using the POET pipeline, and three different phase function models, following methods laid out in previous works as outlined here: (1) we tested the standard two-term sinusoidal phase function with \cite{Mandel2002} transit and eclipse models \cite[POET applications discussed by][among others]{Stevenson2017}; (2) we tested a third-order spherical harmonics phase function \citep[POET applications discussed by][]{Kreidberg2018} using the \texttt{SPIDERMAN} package for the phase and eclipse models \citep{Louden2018} in combination with \texttt{BATMAN} transits models \cite{Kreidberg2015}; and (3) we tested a Lambertian sphere phase model in combination with \citeauthor{Mandel2002} transits and eclipses.
}

\section{Degeneracy between BLISS mapping and PRF decorrelation, as identified in WASP-43b 4.5 micron photometry} \label{degeneracy}
\begin{figure}
	\centering
	\epsscale{1.15}
    \plotone{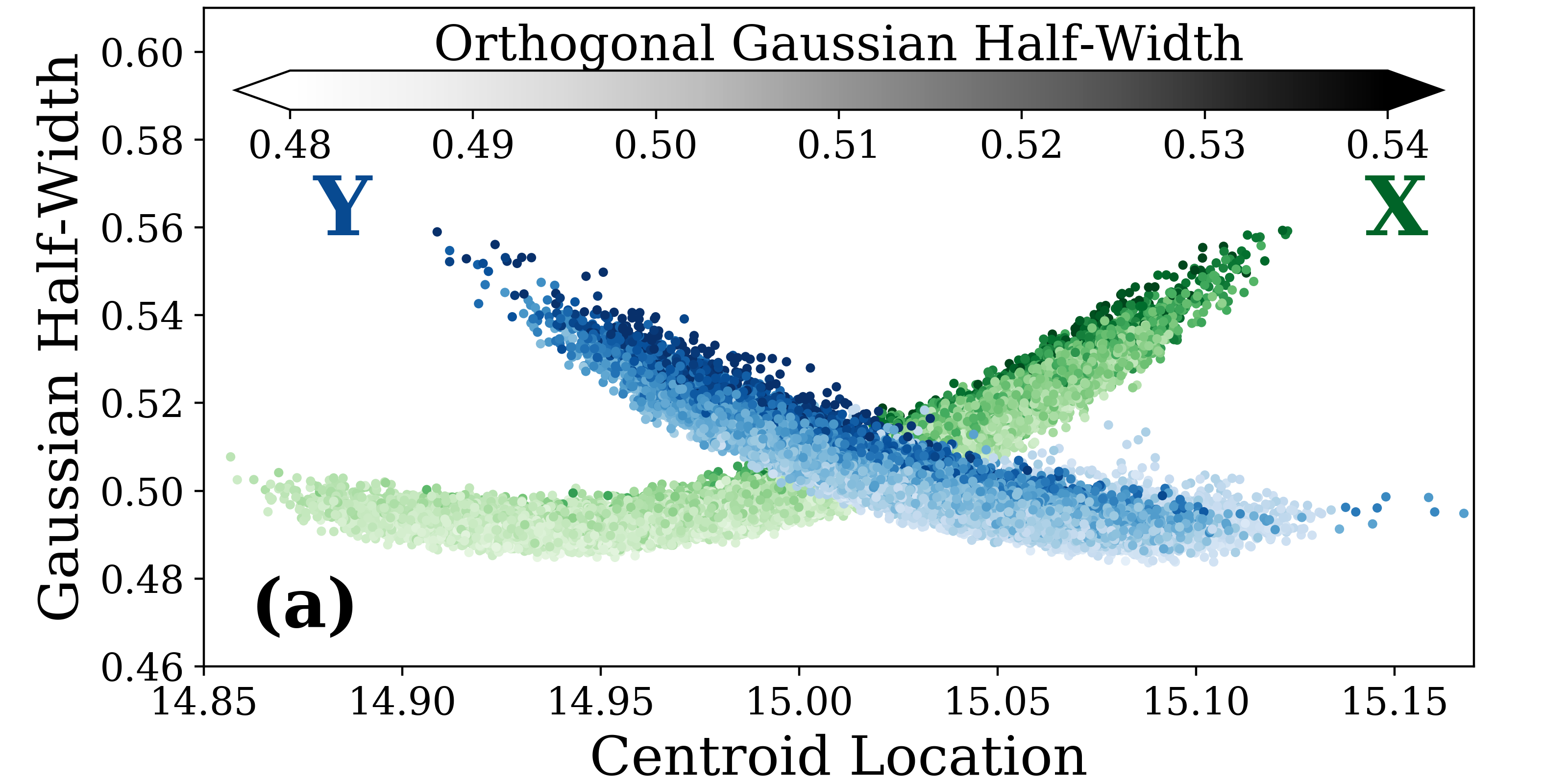}
    \plotone{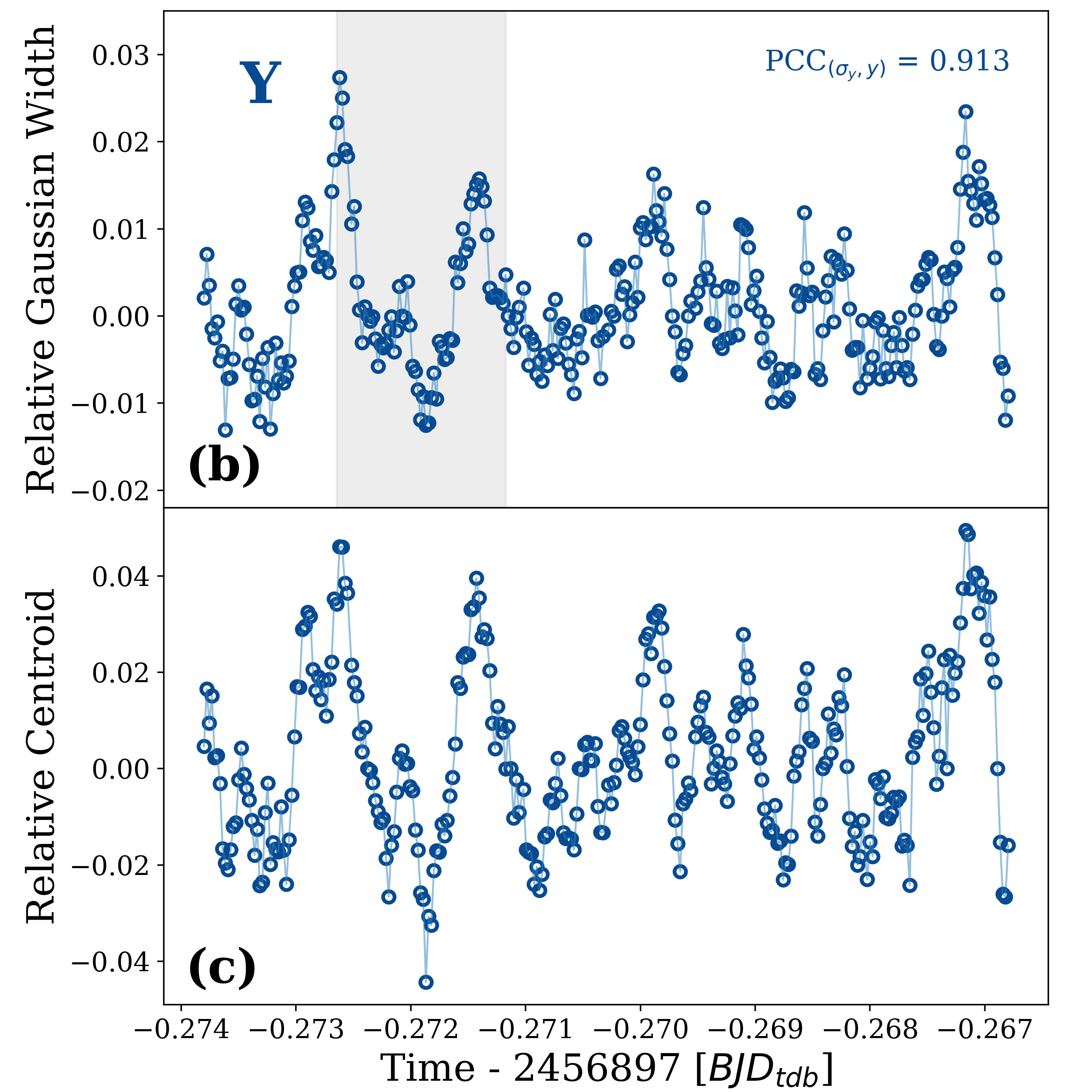}
    \caption{\small{\textbf{Panel (a):} Gaussian half-width vs. centroid for X (green) and Y (blue) for the 4.5$\micron$ WASP-43b phase curve data set. \added{The blue and green shading, as demonstrated by the grey scale colorbar, represents the Gaussian half width in the orthogonal direction, with light values corresponding to smaller values and vice-versa}. \textbf{Panels (b) and (c):} Zoomed in on a random set of 300 exposures, (b) shows the relative Gaussian half-widths and (c) shows the relative centroids in the Y direction. The Pearson Correlation Coefficient for this set of 300 exposures is calculated to be 0.913.}} \label{plt:degen}
\end{figure}
\par As the centroid drifts from the center of an IRAC pixel towards one edge \added{due to {\em Spitzer} pointing motions}, the absolute flux level measured tends to decrease and the shape of the PRF trends towards an oval instead of a circle. BLISS mapping removes the trend in flux by generating an intra-pixel sensitivity map for each data set to explain the differences between phase curve models and raw data, while PRF detrending uses a polynomial function of the Gaussian-width to decorrelate against flux in order to remove the secondary effect of \added{photometry losses due to the mismatch between the circular aperture and} oval PRF \deleted{after the BLISS map variations have been removed}. As we demonstrate in Figure \ref{plt:degen}, the centroids and Gaussian half-widths used in these two detrending techniques are highly correlated. In the top panel we plot the Gaussian half widths in $Y$ (blue) and $X$ (green) as a function of the centroid location, with the shading from light to dark representing the Gaussian half width of the mutually orthogonal axis. The bottom panel shows a zoom-in of the Y-centroids and Y-Gaussian widths over the time scale of approximately 5 frames of 64 exposures each (the grey bar highlights a single group of 64 exposures), from this we clearly see that the BLISS mapping and PRF-FWHM decorrelation techniques are effectively decorrelating against the same parameter twice -- we calculate the Pearson Correlation Coefficient for the Gaussian Width and Centroids in this data set to be 0.725 in the Y-dimension and -0.540 in the X-dimension over the entire set of exposures. Naturally, when two functional forms describe the same effect and are determined by extremely similar inputs, we would expect a strong degeneracy between them if both are used simultaneously to fit observed data. 
\begin{figure}
	\centering
	\epsscale{1.15}
    \plotone{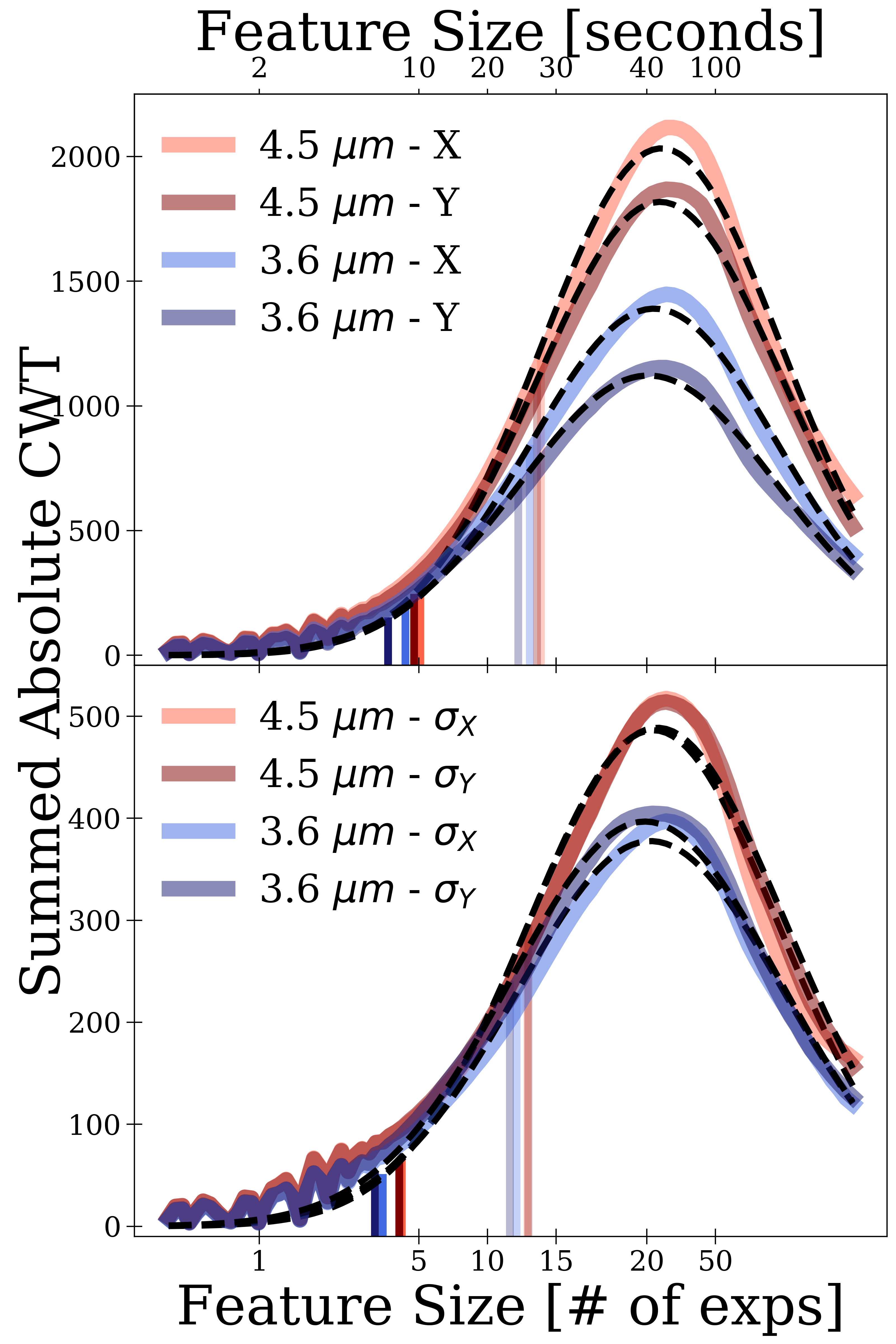}
    \caption{\small{\textbf{Top:} Summed continuous wave transforms for the centroids at both 4.5 $\micron$ (red) and 3.6 $\micron$ (blue). Both the X (lighter color) and Y (darker color) dimensions are shown. \textbf{Bottom:} Same for Gaussian half-widths. For all curves, we fit Gaussians (black dashed lines) to estimate a 2$\sigma$ limit on the scale of repeating features in the data sets, shown by the dark vertical lines. The scale of repeating features is shown in both exposure numbers and seconds.}} \label{plt:scaleogram}
\end{figure}
\par We used a wavelet analysis to further explore the substructure of the centroids and Gaussian widths identified in Figure \ref{plt:degen}. We find that they exist prominently throughout the entire 3.6 and 4.5 micron phase curve data sets for WASP-43b. To demonstrate this, in Figure \ref{plt:scaleogram} we show continuous \replaced{wave}{wavelet} transforms (CWT) for both the centroid and Gaussian widths (in both $X$ and $Y$ dimensions and {\em Spitzer} channels) summed over their entire respective data sets. These summed CWTs demonstrate the relative strength of repeating features of a given timescale. We see that for all curves, repeating features with a size of approximately 20 exposures, or 40 seconds, are most prominent. For each curve, we fit a Gaussian and define the 2$\sigma$ level as the maximum bin size such that no repeating features crucial to the BLISS map fits or PRF-FWHM decorrelation are smoothed over. This corresponds to a bin size of approximately 4 exposures, or 8 seconds in all cases. \added{\cite{Grillmair2012} identified repeating features on the scale of 1-3 minutes in {\em Spitzer} exoplanet observations and attributed them to spacecraft jitter; the dominant 40 second features we identify in these WASP-43b 3.6- and 4.5- $\micron$ phase curves are likely a result of the same phenomenon.}
\begin{figure*}[t]
	\centering
	\epsscale{1.15}
    \plotone{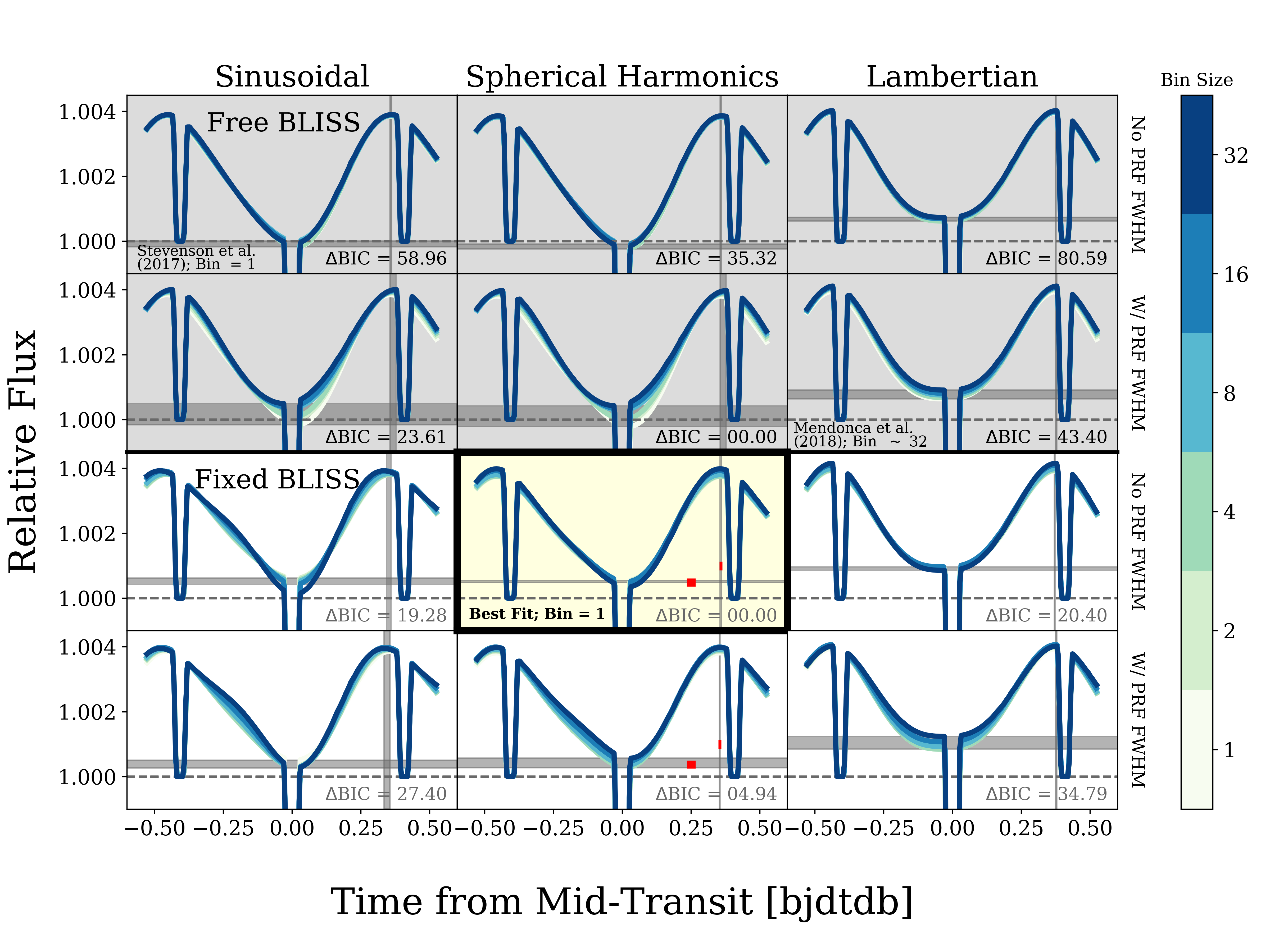}
    \caption{\small{Best fits for the WASP-43b 4.5$\micron$ phase curve. For all rows, the \textbf{left} column represents fits using an asymmetric sinusoidal phase function, the \textbf{center} column represents fits using a third order spherical harmonic phase function, and the \textbf{right} column represents fits using a Lambertian sphere phase function. \textbf{Top 2 Rows:} A free BLISS map that is fit for by the data itself with the \textit{first} row not using PRF-FWHM detrending and the \textit{second} row using PRF-FWHM detrending. \textbf{Bottom 2 Rows}: The BLISS map is held to the fixed sensitivity map we generate in this work with the \textit{first} row not using PRF-FWHM detrending and the \textit{second} row using PRF-FWHM detrending. The different colors correspond to different temporal binning, as represented by the colorbar. When a free BLISS map is used in combination with PRF FWHM, increasing bin size decreases the phase offset and increases the night side flux measured (decreasing the phase amplitude). The combination of functions used by \cite{Stevenson2017} and \cite{Mendonca2018} are labeled, as well as the best fitting set of functions for this work.  The vertical shaded regions correspond to the range of hotspot offsets a given set of models returns, while the horizontal shaded regions correspond to the range of night side emission values a given set of models returns, shown in more detail in Figure \ref{plt:offs_amps}. The best fit panel includes red bars representing the uncertainty on the night side emission and hotspot offset. This set of fits demonstrates the degeneracy of a free BLISS map and PRF-FWHM detrending, with the range of night side emission and offsets being outside the typical uncertainties. \added{In the top two row we include the $\Delta$BIC values amongst the free BLISS fits, while the bottom two rows include the $\Delta$BIC values amongst the fixed BLISS fits, all for the no binning case.}}} \label{plt:bins}
\end{figure*}
\begin{figure}[t]
	\centering
	\epsscale{1.15}
    \plotone{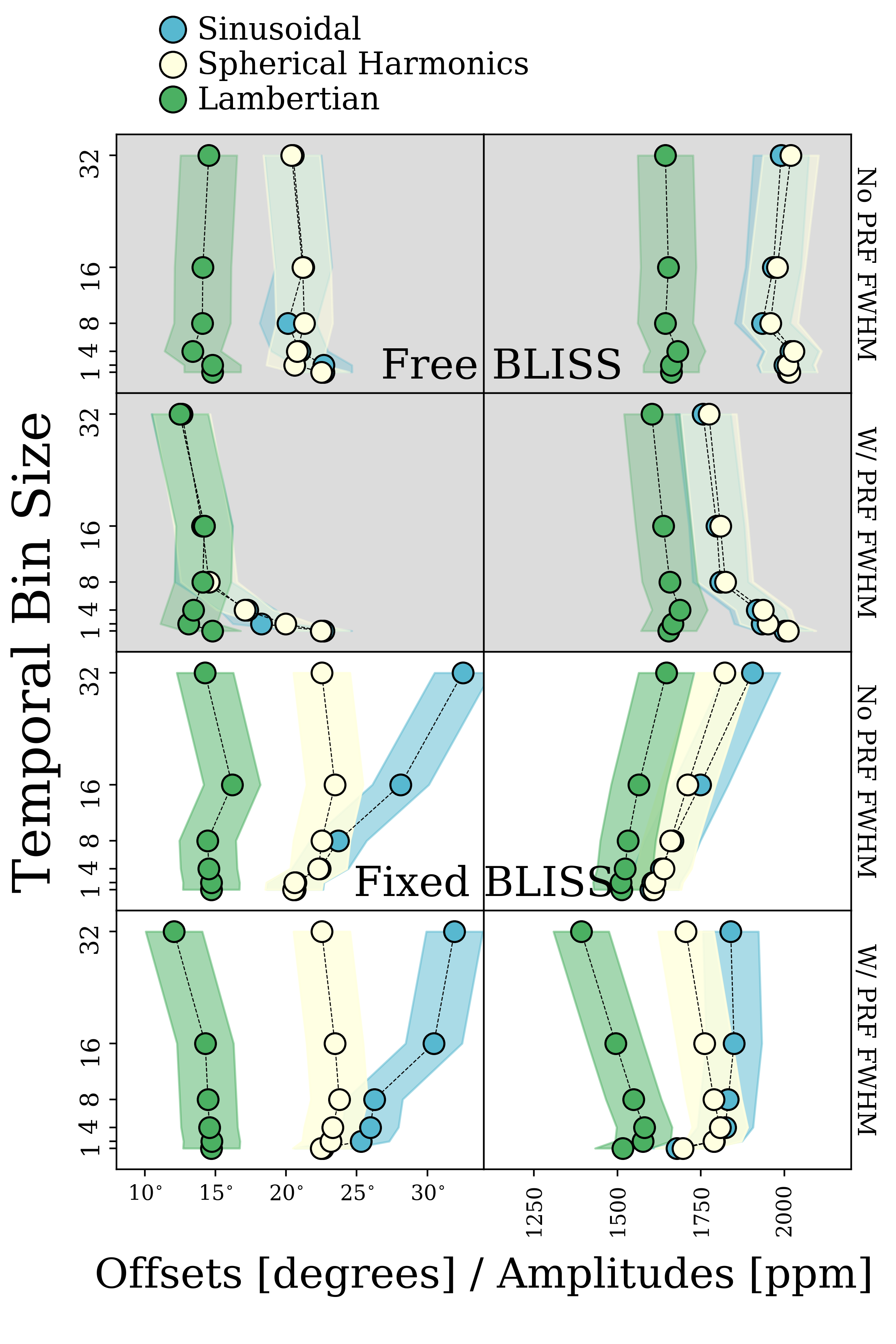}
    \caption{\small{\added{Hotspot offsets and phase amplitudes as a function of bin size, shown for all model combinations presented in Figure \ref{plt:bins}. As before, the top two rows show the use of the free BLISS Map, and the bottom two show the new fixed sensitivity map.}}} \label{plt:offs_amps}
\end{figure}

\par Using the 4.5$\micron$ WASP-43b phase curve data as a test case, we further explore how the resulting BLISS map and PRF-FWHM fits become degenerate when the temporal bin size is larger than the scale of the important substructure identified in Figures \ref{plt:degen} and \ref{plt:scaleogram}, or a bin size of $\sim$4 exposures. \added{In all cases, temporal binning takes place prior to model fitting, including systematic removal.} To demonstrate this, in Figure \ref{plt:bins} we present model fits to the 4.5$\micron$ data of WASP-43b using an asymmetric sinusoidal, third order spherical harmonic, and Lambertian phase model at bin sizes 1, 2, 4, 8, 16, and 32. We also note the model combinations used by \cite{Stevenson2017} (asymmetric sinusoidal, no PRF FWMH, no temporal binning) and \cite{Mendonca2018} (Lambertian, PRF FWHM, temporal bin size of $\sim$ 30) and highlight the model suggested by $\Delta$BIC values for a fixed sensitivity map (third order spherical harmonics, no PRF FWHM, see Section \ref{W43}) when binning is not used. \added{In Figure \ref{plt:offs_amps} we show that increasing temporal binning when using a free BLISS and PRF FWHM detrending results in a smaller phase offset and a larger night side flux (smaller phase amplitude). It is important to note that while it appears that models diverge with the use of the fixed sensitivity map, this is to be expected when the systematic model is held constant and different phase shapes are being fit to the exact same de-trended data. For the fixed sensitivity map cases, the no-binning fits (bin size of 1) have the same phase amplitude and hotspot offset for both the sinusoidal and spherical harmonic options, while the Lambertian is no longer a good fit to the data, also demonstrated by the $\Delta$BIC values for the no-binning fits on Figure \ref{plt:bins}.} The models trend towards over fitting for temporal bin sizes larger than 4 when comparing the expected and measured RMS of the residuals, as shown in Figures 4 and 17 of \cite{Morello2019}. The evidence of over fitting for bin sizes larger than $\sim$4 exposures is in agreement with our suggestion from Figure \ref{plt:scaleogram} that binning above sizes of 4 exposures (or 8 seconds) removes important repeating features.
\par In agreement with \cite{Mendonca2018}, and as suggested by the $\Delta$BIC values of the fits, we find that our data prefers the addition of the PRF-FWHM detrending when a free BLISS model is used. However, when a free BLISS map is used in combination with PRF-FWHM detrending, increasing temporal bin size results in the \deleted{best} fit phase curve drifting away from the no-binning case, resulting in phase amplitudes and offsets that do not agree within uncertainties as the bin size is increased. We attribute this to binning causing the substructure of Figure \ref{plt:degen} to be smoothed over, resulting in the information needed to accurately fit the \replaced{two functions}{map and PRF functions} independently being lost. This degeneracy exists regardless of the underlying phase curve models used, though it is less apparent for a Lambertian phase model, which is a restricted form of the asymmetric sinusoid function.
\par As shown in the \replaced{two non-shaded panels}{bottom two rows of Figure \ref{plt:bins}}, after applying the fixed sensitivity map we present in Section \ref{mastermap}, we find a significant decrease in the variations between \deleted{best} fit models caused by temporally binning the data. Further, the $\Delta$BIC values of the fits no longer suggest that there is anything to gain by adding the PRF-FWHM detrending for the 4.5$\micron$ WASP-43b data set. We address \replaced{our understanding for the lack of a need for the PRF FWHM detrending}{the reason why PRF detrending is unnecessary} when the fixed sensitivity map is used in Section \ref{mastermap}. \added{The single best-fit function, discussed in Section \ref{W43} and highlighted in Figure \ref{plt:bins}, was chosen by comparing the $\Delta$BIC value of the no-binning cases for all studied phase-functions, both with and without the inclusion of PRF-detrending.}
%

\section{A New Fixed Intra-Pixel Sensitivity Map} \label{mastermap}
\par In this work we create a new fixed intrapixel sensitivity map to both address the free BLISS map - PRF FWHM detrending degeneracy, and to eliminate issues when single phase curve observations over multiple AORs have non-overlapping centroids. Our fixed sensitivity map is generated with the 2D Gaussian centroiding routine that is used in the POET pipeline. This map is fully integrated in the POET pipeline such that the BLISS mapping routine will hold the map values constant in regions where the data centroids overlap with the fixed sensitivity map and allows the data to fit itself in regions where no fixed sensitivity map exists. This allows for the map to be applied to data sets that only partially overlap with our fixed sensitivity map.
\par Following \cite{Ingalls2012}, we used the calibration star KF09T1 (TYC 4212-1074-1) in the 3.6$\micron$ channel and BD+67 1044 (NPM1+67.0536) in the 4.5 $\micron$ channel. Many additional AORs with these calibration stars centered in the `sweet spot' have been taken since the \citeauthor{Ingalls2012} map was published, resulting in our maps covering a wider region of the pixel with a total of 356 and 240 AORs at 3.6 and 4.5 $\micron$, respectively for a total of 1,636,697 exposures at 3.6$\micron$ and 3,712,830 exposures at 4.5$\micron$. In total, the calibration data set we analyze contains 3.6$\times$ more 3.6$\micron$ data and 12.9$\times$ more 4.5$\micron$ data than \cite{Ingalls2012}. \added{\cite{Ingalls2018} update their map using nearly the same data set as we use here, but temporally bin the frames by a factor of 4 prior to performing their aperture photometry. While this map contains the same total number of observations, the different centroiding method and temporal binning warrant our generation of our new fixed sensitivity map in this work.}
\begin{figure}
	\centering
	\epsscale{1.15}
    \plotone{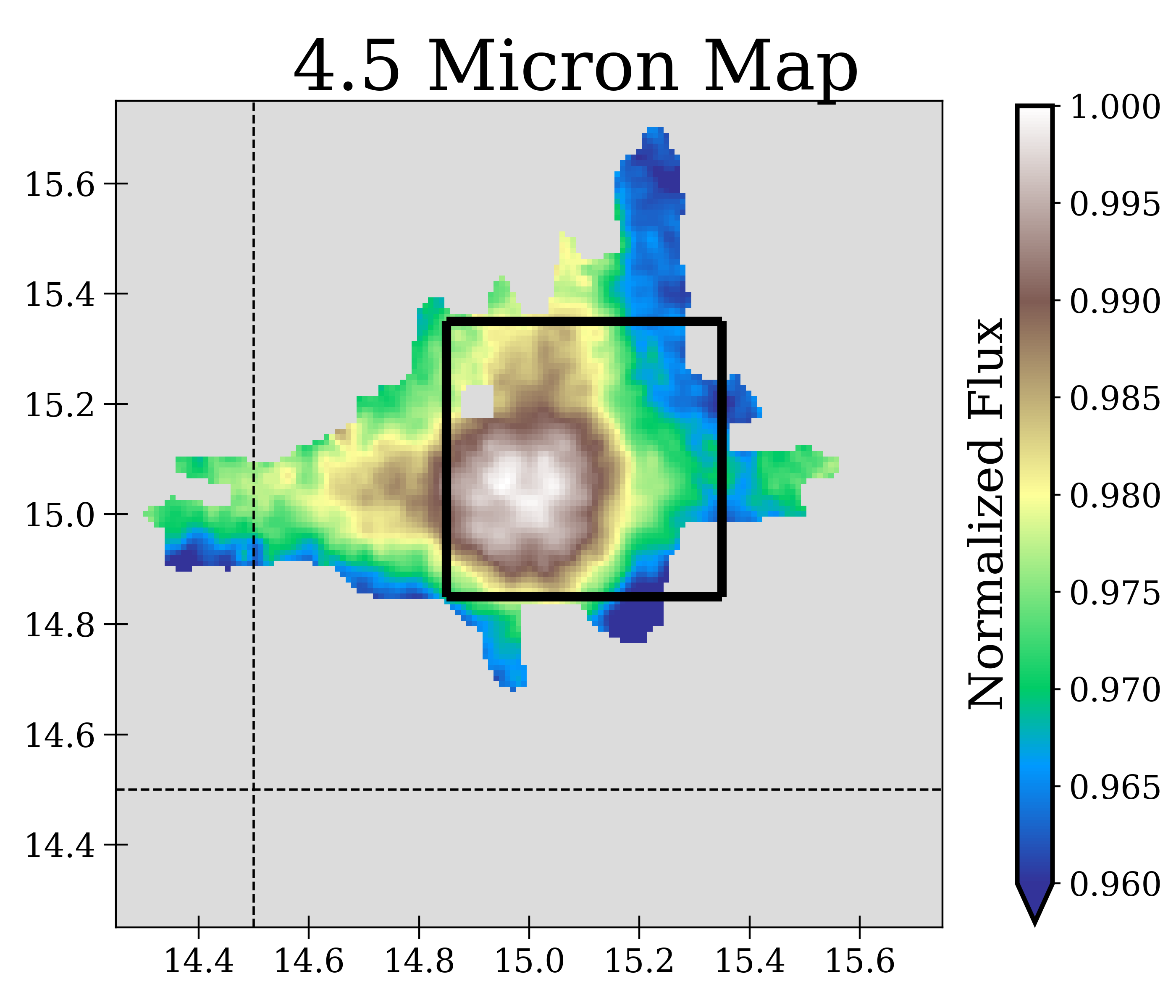}
    \caption{\small{Our final fixed sensitivity map at 4.5$\micron$, following the steps outlined in Section \ref{sec:method}. The black box represents the pixel `sweet spot' defined by \cite{Ingalls2012}.}} \label{plt:map}
\end{figure}
\subsection{Method} \label{sec:method}
\par In order to generate a intrapixel sensitivity map that is applicable to the POET pipeline, we used the centroiding and photometry routines of the POET pipeline to generate centroids \{$x, y$\} and fluxes \{$F$\} for all exposures of the above objects in their respective channels. The rest of the process we used to create the intrapixel sensitivity map is outlined below:
\begin{enumerate}
\item We identified groups of AORs that were taken consecutively and combined them into time-grouped sets of centroid and flux values. 
\item We mapped each time-grouped set of \{$x, y, F$\} values onto a grid of points with 0.01 pixel spacing, requiring a minimum of 2 points in a grid point, while taking the median \{F\} within each grid point as the `true' flux at that grid location. 
\item To account for differences in instrument calibrations, we took each time-grouped binned map and rescaled it relative to the rest of the map. This was done iteratively until the rescale factors for all time-grouped binned maps converged such that they vary by less than 10 ppm in a given iteration. The rescale factors calculated in this step are applied to the original non-grid mapped sets of \{$x, y, F$\} values.
\item We next applied a bicubic spline fit to the resulting map. This step first maps the entire set of rescaled \{$x, y, F$\} pairs onto a grid of points with grid sizes of 0.001 of a pixel, with no requirement on the minimum number of points in a grid. We then defined a grid of knots for the spline fit at a scale of 0.02 of a pixel (1 knot per 5 data grid points). The bicubic spline fit was then evaluated on the original 0.001 pixels/point grid to generate a smoothed high resolution intrapixel sensitivity map. 
\item We removed isolated data grid points from the fit in an iterative process to include groups of 3 or less grid points more than 1 knot step away from any other points.
\item Finally, we interpolated over gaps in the high resolution smoothed map less that were less than 1 knot step wide and then removed the outer 3 rows of the high resolution smoothed map to eliminate boundary effects of the spline fit.
\end{enumerate}
\par The selection of the grid sizes was done to minimize the reduced $\chi^2$ of the spline fit compared to the raw data and results in a map that is five times higher resolution than the typical spatial binning of data in POET for BLISS map fitting. The final map from step 6 is fully integrated into POET so that data that overlaps the fixed sensitivity map is calibrated by the map, while non-overlapping data is allowed to self calibrate. We find no discontinuities when the BLISS mapping routine shifts from fixed-values to free-values in cases where the set of centroids only partially overlap with our fixed sensitivity map.
\par Because our fixed sensitivity map is initially reduced using the same methods we use for phase curve data reduction, the fluxes measured for the calibration frames encapsulate both the intrapixel sensitivity variations and the flux variations due to the changing PRF FWHM as the centroid approaches the edge of the pixel. We find that the use of the fixed sensitivity map sufficiently removes these variations in the 4.5$\micron$ WASP-43b phase curve data set such that PRF FWHM detrending is no longer suggested -- i.e. when examining the residual fluxes as a function of the PRF FWHM after the fixed sensitivity map variations are removed, the best fit is simply a constant value instead of a first or second order function of the PRF FWHM, as is used in \cite{Mendonca2018}. 
\par Our new fixed sensitivity map is highly sensitive to the aperture size used when performing the aperture photometry, and phase curve data sets analyzed with the map require one made with the same aperture size as the data in order to generate reasonable fits. Here we focus on our fixed sensitivity maps generated at an aperture radius of 2.0 pixels for the 4.5$\micron$ channel and 3.0 pixels for the 3.6$\micron$ channel with the background subtraction from 7.0 to 15.0 pixels to directly relate to the WASP-43b data we re-analyzed, but maps of varying aperture sizes between 2.0 and 4.0 in quarter-pixel steps are available with the POET pipeline. Figure \ref{plt:map} presents our map for 4.5$\micron$, while Section \ref{sec:var} discusses the time variability we found at 3.6$\micron$.
\subsection{Time Variability in the 3.6 $\mu$m Channel} \label{sec:var}
\par We found that the 3.6$\micron$ calibration AORs exploring the sweet spot for the calibration star KF09T1 were difficult to rescale via step 3 of our method, and the resulting map did not fit the WASP-43b 3.6$\micron$ phase curves well. As time variability of the instrument systematics has been a concern for {\em Spitzer} IRAC, we explored how that may be affecting both the 3.6 and 4.6 $\micron$ calibration data sets identified in the previous section.
\begin{figure*}[h!]
	\centering
	\epsscale{1.05}
    \plotone{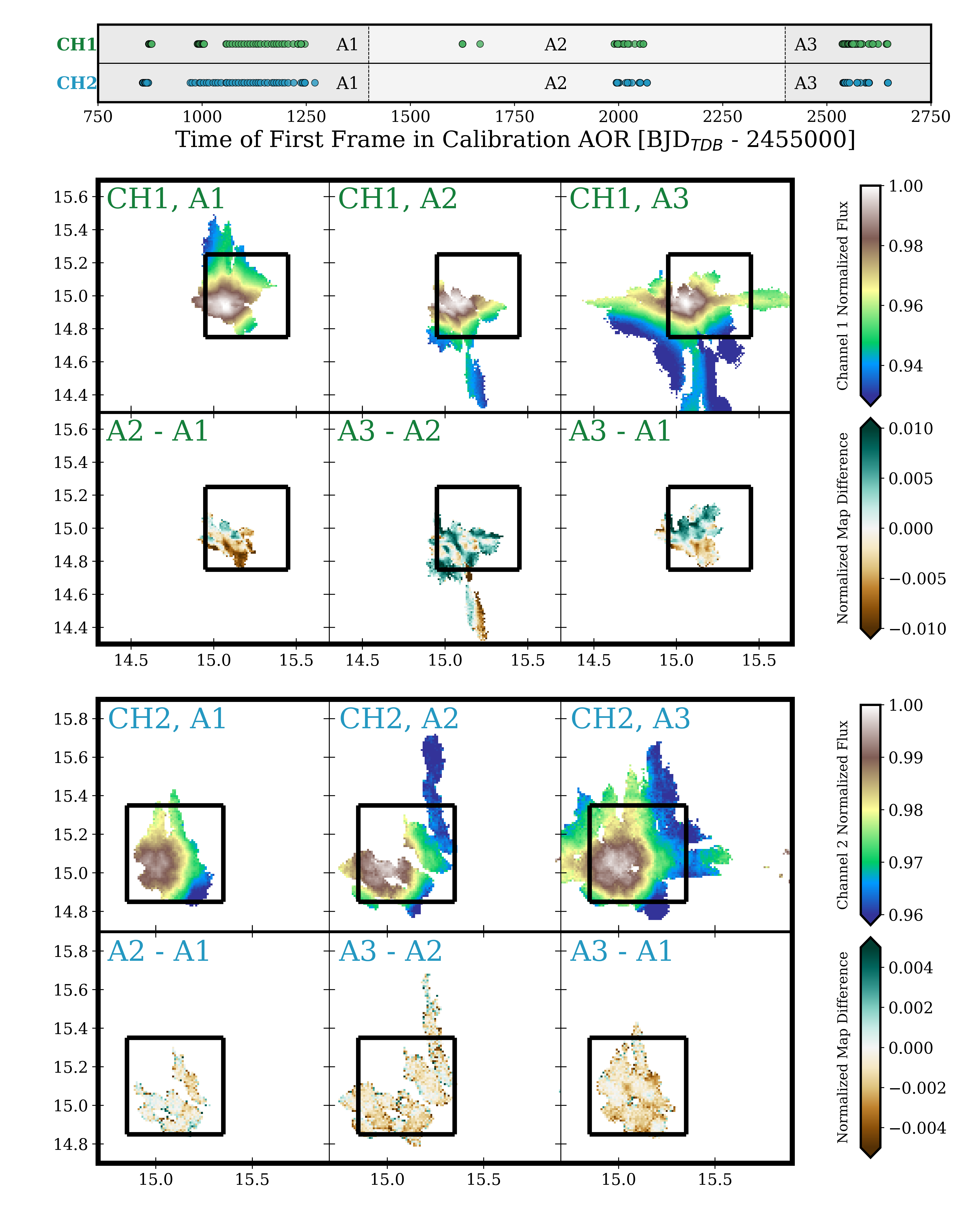}
    \caption{\small{Here we demonstrate that the 3.6$\micron$ calibration data is time variable. \textbf{Top:} times for all calibration AORs we use in both channels. We identify 3 discrete `groups' and subdivide the calibration data based on these definitions. \textbf{Middle:} \textit{Upper Middle:} Binned centroids and fluxes for all three time groups for 3.6$\micron$ and \textit{Lower Middle:} Differences between the discrete time groupings. \textbf{Bottom:} Same, but for 4.5$\micron$. Note that for these difference maps, the scaling on the colorbar for 3.6$\micron$ is twice that of 4.5$\micron$.\added{In all panels, the black box denotes the commonly identified pixel ``sweet spot''.}}} \label{plt:timevar}
\end{figure*}
\par To explore the time variability of both channels, in Figure \ref{plt:timevar} we show the map from step 2 of our method in our identified discrete time groups A1, A2 and A3 for both channels. The 3.6$\micron$ channel is presented in the first set of maps, while 4.5$\micron$ is presented in the second set of maps. There is clear structure in the difference maps between the discrete time groups at 3.6$\micron$, while 4.5$\micron$ is mostly well distributed scatter around zero. With this we conclude that either the sensitivity of the 3.6$\micron$ channel is time variable or the calibration star KF09T1 itself is in fact variable. The 4.5$\micron$ channel exhibits no significant variability in time.
\par The {\em Spitzer} data archive shows three other calibration stars that were used to explore a small region of the 3.6$\micron$ sweet spot, KF03T2, KF06T1, and KF06T2. However, the data for these stars do not cover enough of the pixel or a sufficiently long time baseline to determine whether the time variation we see in KF09T1 is due to the instrument or the star itself. Therefore, we do not suggest the use of a fixed sensitivity map at 3.6$\micron$ and focus only on the applications of the 4.5$\micron$ map to WASP-43b.

\section{A Reanalysis of the WASP-43\lowercase{b} phase curves} \label{W43}
\par WASP-43b \citep[discovered by][]{Hellier2011} is a 1.78 $\pm$ 0.1 M$_J$ and 0.93$^{+0.07}_{-0.09}$ R$_{J}$ exoplanet on a 19.5 hour orbit around a K star. It is particularly well suited to phase curve observations due to its relatively short orbital period and temperature ratio with the host star. 
\par \cite{Stevenson2017} presented {\em Spitzer} phase curves at 3.6 and 4.5 $\micron$ \deleted{(Programs 10169 and 11001, PI: Kevin Stevenson)}. In addition to the transit and eclipse models, they used a best fit model of a two-term sinusoidal phase function with varying offsets to allow for asymmetry in the phase curve, a linear ramp, and the BLISS mapping decorrelation method. They further found a significant difference between the two 3.6$\micron$ visits resulting in night side emission during visit one and no night side emission at 3.6$\micron$ during visit two. They did not detect night side emission at 4.5$\micron$. Due to nonphysical model results and non-overlapping AORs in the first visit, they suggested that the second visit at 3.6$\micron$ is more accurate and report no night-side emission in either channel for WASP-43b with an upper limit of 650 K for the night side temperature at 4.5$\micron$. They reported a phase offset of 21.1$^{\circ}$ $\pm$ 1.8$^{\circ}$.
\par \cite{Mendonca2018} reanalyzed the above WASP-43b phase curves using, in addition to the transit and eclipse models, a Lambertian phase function that is by definition a symmetric sinusoid, a constant ramp, and the BLISS mapping and PRF-FWHM decorrelation methods. From the reported BIC values, we infer that they used a bin size of ~32 for the analysis, above the value we identified as resulting in degenerate values between the BLISS map and PRF-FWHM functions. They did not find the same difference between the two 3.6$\micron$ observations as \cite{Stevenson2017}, and both 3.6 and 4.5 $\micron$ results show measurable night-side emission for WASP-43b. The authors only reported their phase curve amplitudes, so we assume equal error contribution from the day and night side fluxes. Taking the minimum flux level from their best fit phase curve, we used this assumption to calculate a night side temperature and its uncertainty based on their analysis. Using the \citeauthor{Mendonca2018} best fit, we compute a night side disk integrated temperature of 913 $\pm$ 55 K at 4.5$\micron$, 4.8$\sigma$ higher than the upper limit imposed by \cite{Stevenson2017}. The phase offset of this analysis is measured at 12$^{\circ}$ $\pm$ 3$^{\circ}$, a 2.6$\sigma$ difference from \cite{Stevenson2017}.
\par The discrepancy between these two results inspired this work to reanalyze WASP-43b using the POET pipeline with the PRF-FWHM decorrelation method, during which we discovered the degeneracy between the two decorrelation methods. While this analysis was in preparation, \cite{Morello2019} published a third analysis of WASP-43b using Independent Component Analysis (ICA) to detrend the data instead of physically defined models. The phase curve model is an asymmetric sinusoid and the data were binned to a size of 8 exposures for the analysis. Their best fit models show a significant difference between the two 3.6$\micron$ observations \citep[as in][]{Stevenson2017} while also detecting night-side emission at both 3.6 and 4.5 $\micron$. They reported a disk integrated night side temperature of 700$^{+68}_{-93}$ K at 4.5$\micron$, 2.3$\sigma$ lower than \cite{Mendonca2018} but within 0.5$\sigma$ of the upper limit of \cite{Stevenson2017}. 
\subsection{WASP-43b with a fixed BLISS map}
\begin{figure*}[t!]
	\centering
	\epsscale{1.1}
    \plotone{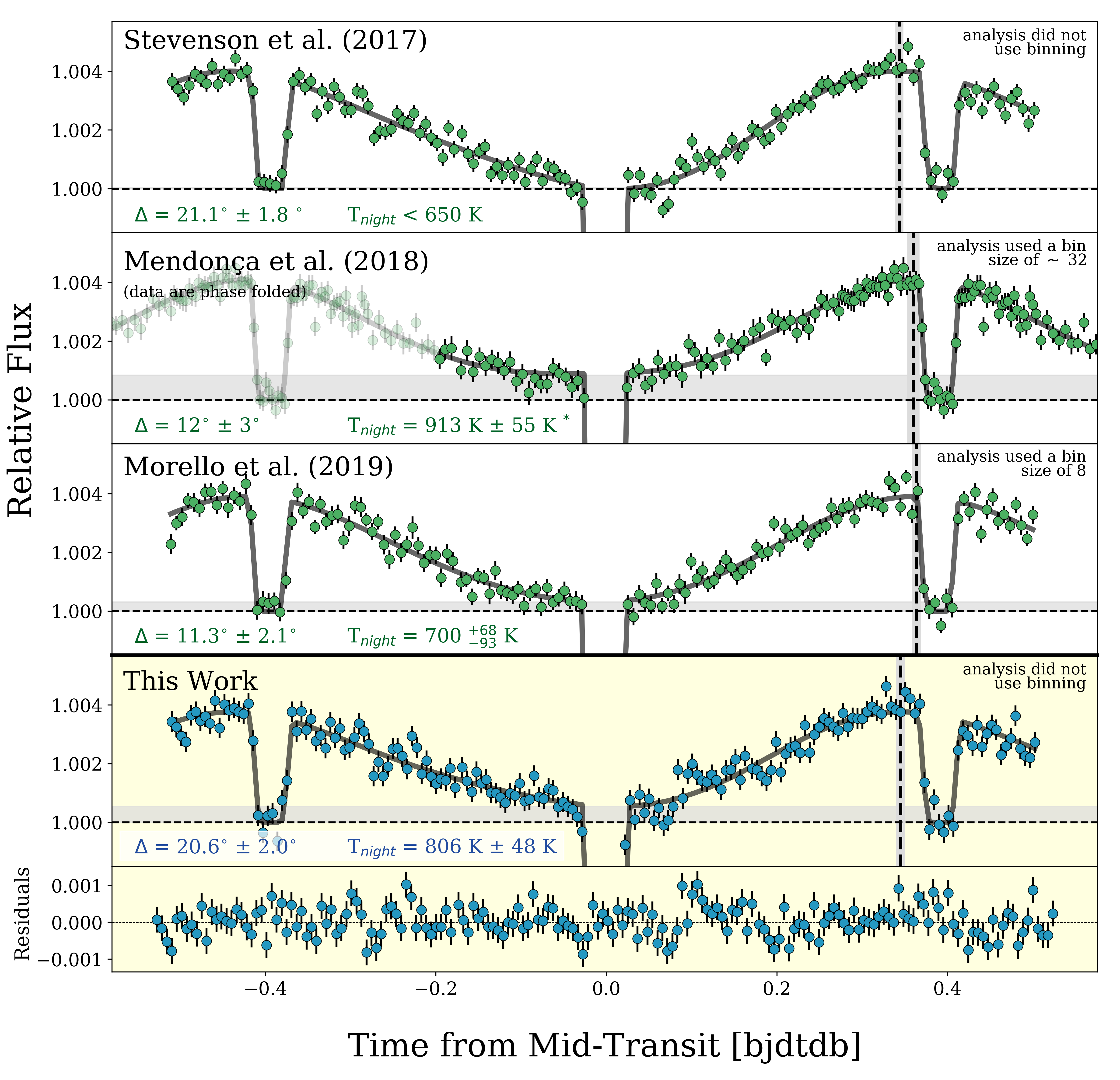}
    \caption{\small{Best fit data and models from \cite{Stevenson2017}, \cite{Morello2019}, \cite{Mendonca2018}, and this work. The dashed vertical lines in each panel correspond to the location of peak flux with a shaded region for the 1$\sigma$ uncertainties. The dashed horizontal lines correspond to the stellar flux level with the shaded region corresponding to the level of night side emission for each analysis. The results of the night side emission presented in this work are in agreement with \cite{Mendonca2018} within 1.5$\sigma$ and \cite{Morello2019} within 1.3$\sigma$. The hotspot offset agrees with \cite{Stevenson2017} to within 0.19$\sigma$.}} \label{plt:W43}
\end{figure*}
\begin{table*}
    \centering
    \begin{tabular}{|l|c|c|c|c|c|c|} 
    \hline
        {  }    &   Hotspot         &   Amplitude   &    \multicolumn{2}{|c|}{F$_p$/F$_S$ [ppm]}  &   \multicolumn{2}{|c|}{Temperature [K]}   \\ \cline{4-7}
        { }     &   Offset [$^{\circ}$] &   [ppm]     &  Dayside     &   Nightside       & Dayside        & Nightside                \\ \hline \hline
        \cite{Stevenson2017}& 21.1 $\pm$ 1.8  &   1995 $\pm$ 70   &   \textit{[4008]} &    \textit{[19]}    &   1512 $\pm$ 25   & \textless650    \\
        \cite{Mendonca2018} & 12 $\pm$ 3    &   1629 $\pm$ 125  &   \textit{[4098]} &     \textit{[841]}   &   \textit{[1578]} & \textit{[913]}  \\
        \cite{Morello2019}  &   11.3 $\pm$ 2.1 &  \textit{[1800]}  &   3900 $\pm$ 120  &   300 $\pm$ 150   &   1502 $\pm$ 18   &   700$^{+68}_{-93}$   \\
        This Work   &   20.6 $\pm$ 2.0  &   1613 $\pm$ 83   &   3768 $\pm$ 116 &    542 $\pm$ 124   & 1485 $\pm$ 41 &   806 $\pm$ 48 \\ \hline 
    \end{tabular}
    \caption{Best fit values for the 4.5 $\micron$ WASP-43b phase curve as reported in their respective papers.\added{Numbers in brackets were not originally reported in the respective papers, but were calculated here from best-fit phase functions provided by the authors.}} 
    \label{table:bestfit} 
\end{table*}
\par To address the discrepancies in the three previous analyses, we apply our new fixed sensitivity map, and fit the 4.5$\micron$ WASP-43b with the BLISS map fixed while testing to ensure the results are robust against binning. We test constant, linear, and quadratic ramps as well as an asymmetric sinusoidal, 3rd order spherical harmonic, and Lambertian phase functions. For all of the above we also fit the data both with and without the PRF-FWHM decorrelation model. We find no strong correlation between fit model parameters and bin size when the BLISS map is held to the fixed sensitivity map values. The results of this model combination exploration is summarized in Section \ref{degeneracy} and Figure \ref{plt:bins}. At 4.5$\micron$ we find the best fit model combination is a 3$^{rd}$ order spherical harmonic with a linear ramp and no PRF-FWHM decorrelation based on the $\Delta$ BIC values. We note that the sinusoidal phase function gives us the same amplitude and phase offset results as the third order spherical harmonics, and that the Lambertian phase function no longer matches the data well when the fixed sensitivity map is used. Figure \ref{plt:W43} summarizes the reduced data and respective best fits for all previous analyses of this 4.5$\micron$ phase curve, as well as the results of this work with our new fixed sensitivity map. We include phase offsets and night side temperatures for all results, and note the temporal binning used for each analysis method. 
\par With this new 4.5$\micron$ analysis, we measure a night side planet-to-star flux ratio of 542 $\pm$ 124 ppm and a day side planet-to-star flux ratio of 3768 $\pm$ 116 ppm, corresponding to a phase curve amplitude of 1613 $\pm$ 83 ppm. Using a stellar temperature of 4520 $\pm$ 120 K and log$g$ of 4.645 $\pm$ 0.02 cm/s$^{2}$ \citep{Gillon2012} we derive disk integrated night side and day side brightness temperatures of 806 $\pm$ 48 K and 1485 $\pm$ 41 K, respectively at 4.5$\micron$. Our night side emission levels are within 1.5$\sigma$ of \cite{Mendonca2018} and 1.3$\sigma$ of \cite{Morello2019}. Table \ref{table:bestfit} summarizes the best fit phase curve values at 4.5 $\micron$ as presented in the respective papers.
%
%
\section{Conclusions} \label{conclusions}
\par We have presented a new systematic removal method for high precision 4.5$\micron$ {\em Spitzer} IRAC photometry with specific applications to the dozens of unanalyzed, and previously analyzed, {\em Spitzer} phase curves that were observed prior to the end of the mission. \added{The use of this new method will ensure a uniform application of systematics for all phase curve data sets}. The fixed sensitivity map presented is applicable to 2D Gaussian centroiding and is suggested to be used in combination with the POET pipeline for best results. We have additionally presented evidence for temporal variability of IRAC's 3.6$\micron$ channel sensitivity.
\par Our new method results in a 4.5$\micron$ phase curve amplitude that agrees more closely with \cite{Mendonca2018} and \cite{Morello2019} than with \cite{Stevenson2017}, while our hotspot offset is in agreement with \cite{Stevenson2017}. We note that the hotspot offset are not in close agreement with \cite{Mendonca2018} and \cite{Morello2019}, but suggest that this is likely due to the binning used in the \citeauthor{Mendonca2018} and \citeauthor{Morello2019} analyses, as we've shown that binning the data in time affects the best fit light curves, specifically by decreasing the phase amplitude and offset values.
\par Although our results might suggest that a free BLISS map in combination with PRF-FWHM detrending and large temporal binning produce results in line with our fixed sensitivity map fits, \replaced{we caution that this may not hold true for all phase curve data sets}{as of writing there is no evidence that this holds true for all phase curve data sets}. Further, we find that the free BLISS with PRF-FWHM detrending begins to overfit the data for bin sizes larger than 4, suggesting that we cannot put much weight onto the results of such fits. Therefore, we strongly suggest the use of our 4.5$\micron$ fixed sensitivity map to ensure consistent analyses across groups for better comparative exoplanet studies and recommend all further analyses of {\em Spitzer} phase curves be done without temporal binning to ensure that both the data are not being over fit, and an accurate measurement of the phase offset and nightside emission is obtained.

\clearpage
\software{\\ Astropy \citep{astropy,astropy2},
\\ batman \citep{Kreidberg2015},
\\ IPython \citep{ipython},
\\ Matplotlib \citep{matplotlib},
\\ NumPy \citep{numpy},
\\ PyWavelets \citep{Lee2019},
\\ SciPy \citep{scipy},
\\ spiderman \citep{Louden2018},}

\bibliography{REFS}{}
\bibliographystyle{aasjournal}
\listofchanges
\end{document}